%% file: main.tex
\documentclass[runningheads]{llncs}
\input{packages}
\input{commands}

\begin{document}
\title{Balancing Wind and Batteries: Towards Predictive Verification of Smart Grids
\thanks{%
This research has been partially funded by NWO grants NWO OCENW.KLEIN.187 and NWA.1160.18.238, and NWO VENI grant no.\ 639.021.754.
}
}
\titlerunning{Towards Predictive Verification of Smart Grids}

\author{%
Thom S. Badings\inst{1}\,\orcidID{0000-0002-5235-1967}
\and Arnd Hartmanns\inst{2}\,\orcidID{0000-0003-3268-8674}
\and Nils Jansen\inst{1}
\and Marnix Suilen\inst{1}%
}
\authorrunning{T.S.\ Badings, A.\ Hartmanns, N.\ Jansen, and M.\ Suilen}
\institute{Department of Software Science, Radboud University, Nijmegen, The Netherlands \and
University of Twente, Enschede, The Netherlands}
\maketitle
\vspace{-0.4cm}
\begin{abstract}
We study a smart grid with wind power and battery storage.
Traditionally, day-ahead planning aims to balance demand and wind power, yet actual wind conditions often deviate from forecasts.
Short-term flexibility in storage and generation fills potential gaps, planned on a minutes time scale for 30-60 minute horizons.
Finding the optimal flexibility deployment requires solving a semi-infinite non-convex stochastic program, which is generally intractable to do exactly.
Previous approaches rely on sampling, yet such critical problems call for rigorous approaches with stronger guarantees.
Our method employs probabilistic model checking techniques.
First, we cast the problem as a continuous-space Markov decision process with discretized control, for which an optimal deployment strategy minimizes the expected grid frequency deviation.
To mitigate state space explosion, we exploit specific structural properties of the model to implement an iterative exploration method that reuses pre-computed values as wind data is updated.
Our experiments show the method's feasibility and versatility across grid configurations and time scales.
\end{abstract}

\input{1_introduction}
\input{0_preliminaries}
\input{2_power_system_model}
\input{3_discrete_MDP}
\input{4_numerical_study}
\input{5_conclusion_discussion}

\newpage

\bibliographystyle{splncs04}
\bibliography{literature}

\end{document}

%% file: packages.tex
\usepackage{amsmath,amssymb,amsfonts}       
\usepackage{mathtools}

\usepackage[usenames]{color}
\usepackage{xcolor}
\usepackage{graphicx}
\usepackage{xspace}
\usepackage{microtype}
\usepackage{todonotes}
\usepackage{colonequals}
\usepackage[noadjust,nospace]{cite}
\usetikzlibrary{patterns,arrows,arrows.meta,calc,shapes,shadows,decorations.pathmorphing,decorations.pathreplacing,automata,shapes.multipart,positioning,shapes.geometric,fit,circuits,trees,shapes.gates.logic.US,fit}
\usetikzlibrary{backgrounds,scopes}
\usepackage{multirow}
\usepackage{multicol}

\usepackage{caption}
\usepackage{subcaption}
\usepackage{url}
\usepackage{appendix}
\usepackage[pdfpagelabels=true]{hyperref}
\usepackage{cleveref}
\usepackage{tikz} 
\usepackage{pgfplots}
\usepgfplotslibrary{external}
\pgfplotsset{compat=1.8}
\usepackage{mdframed}
\usepackage{enumitem}

\setitemize{noitemsep,topsep=0pt,parsep=0pt,partopsep=0pt,leftmargin=12.0pt}
\setenumerate{noitemsep,topsep=0pt,parsep=0pt,partopsep=0pt,leftmargin=12.5pt}
\setdescription{noitemsep,topsep=0pt,parsep=0pt,partopsep=0pt,leftmargin=12.5pt}

\makeatletter%
\g@addto@macro\normalsize{%
  \setlength\abovedisplayskip{3pt}%
  \setlength\belowdisplayskip{3pt}%
  \setlength\abovedisplayshortskip{-3pt}%
  \setlength\belowdisplayshortskip{3pt}%
}%
\makeatother

\makeatletter
\def\orcidID#1{\smash{\href{http://orcid.org/#1}{\protect\raisebox{-1.25pt}{\protect\includegraphics{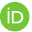}}}}}
\makeatother

\usetikzlibrary{arrows,decorations.pathmorphing,positioning,fit,trees,shapes,shadows,automata,calc} 

%% file: commands.tex
\newcommand{\ie}{i.e.\@\xspace}
\newcommand{\wrt}{w.r.t.\@\xspace}
\newcommand{\eg}{e.g.\@\xspace}

\newcommand{\R}{\mathbb{R}}

\newcommand{\distr}[1]{\ensuremath{\mathit{Dist(#1)}}}

\newcommand{\Finally}{\lozenge \, }

\newcommand{\States}{S}

\newcommand{\Actions}{A}
\newcommand{\initState}{\ensuremath{s_I}}
\newcommand{\transfunc}{T}

\newcommand{\costfunc}{\ensuremath{c}}
\newcommand{\MDP}{\ensuremath{(\States,\Actions,\initState,\transfunc,\costfunc)}}
\newcommand{\mdp}{\ensuremath{\mathcal{M}}}

\newcommand{\ec}{\ensuremath{\mathrm{EC}}}
\newcommand{\expCost}[2]{\ensuremath{\ec^{#1}(\Finally #2)}}

\newcommand{\strat}{\ensuremath{\sigma}}

%% file: 1_introduction.tex
\section{Introduction}\label{sec:introduction}

Electricity grids need to constantly maintain a balance between power supply and demand; imbalances result in frequency deviations, which ultimately lead to critical events like blackouts~\cite{Machowski2006PowerControl}.
The increasing deployment of renewable energy sources such as solar and wind power---which react sharply to hard-to-predict weather conditions---makes maintaining the balance increasingly difficult.
In day-to-day operation, the balancing is managed at two time scales.
One day ahead, the transmission system operator (TSO) schedules conventional generators to match the predicted demand minus the expected renewable generation based on weather forecasts.
During the day, the TSO fills potential gaps introduced by any mismatch between forecast and actual weather conditions by \emph{ancillary services}, which run on a short-term schedule that is updated every few minutes.

On the supply side, the most prominent ancillary service are \emph{spinning reserves} from generators.
They traditionally compensate for contingencies (such as generator failures) and deviations from the predicted demand.
To free capacity for spinning reserves, some generators operate below their rated capacity, making them a costly service.
The TSO's day-ahead plan thus makes a tradeoff between allocating sufficient reserves to mitigate any potential imbalance and minimizing unused generator capacity~\cite{rostampour2017distributedTPS}.
\emph{Demand-side flexibility}~\cite{Badings2019buildings}, on the other hand, is provided by various assets connected to the grid, including batteries~\cite{Kempton2008} and HVAC systems~\cite{chertkov2017ensemble,DBLP:journals/tsg/TahaGDPL19}.
They can reduce or increase the overall power consumption at some time point by injecting or withdrawing electricity into or from the grid.
Such flexibility-based services \emph{shift} power consumption in time rather than changing the total~\cite{MacDougall2013}.
Examples today include the ODFM service in the United Kingdom~\cite{NGESO2020} and various applications of \emph{demand response} throughout the world \cite{Aghaei2013}.
In this paper, we approach the fundamental challenge of short-term scheduling for flexibility-based ancillary~services:
\begin{quote}
Given a power grid with significant uncertain wind power generation,
optimally schedule the deployment of the available ancillary services over a finite horizon
to minimize the expected total grid frequency deviation
without violating any hard constraints on grid stability and operation.
\end{quote}
Hard constraints include a maximum frequency deviation (we use $\pm\,0.1\,$Hz) as well as all generation, transmission, battery, and ramping capacities.
Repeating the optimization every few minutes with new wind measurements leads to a model predictive control (MPC) loop covering a full day of short-term scheduling.

The task can be expressed as a semi-infinite nonconvex stochastic optimization problem, with the potential deviation of wind conditions from the forecast given by some stochastic process.
Already a finite version of this problem is NP-hard and infeasible to solve in practice~\cite{DBLP:journals/mp/MurtyK87,DBLP:books/cu/BV2014}.
Previous work instantiated the stochastic process by a black-box discrete-time Markov chain (DTMC) and then resorted to a sampling-based \emph{scenario optimization} approach~\cite{RostampourBadings2020,DBLP:journals/tac/MargellosGL14}, which linearizes the nonconvex constraints and solves the resulting linear (but still semi-infinite) stochastic optimization problem up to some statistical confidence and error.
The drawbacks of this approach are the \emph{approximation error} introduced by the linearization step and the \emph{statistical error} due to the use of sampling~\cite{DBLP:journals/siamjo/CampiG08}.

\paragraph{Our contribution.}
To overcome the need for both sampling and linearization, we model the problem as a Markov decision process (MDP)~\cite{DBLP:books/wi/Puterman94}. 
MDPs combine probabilistic choices, which we use to follow a white-box DTMC for the wind errors, and nondeterminism, which we use to capture the service deployment decisions to optimize over.
A direct cast of the problem into an MDP would yield continuous state and action spaces:
state variables would represent continuous quantities (\eg grid frequency), and control decisions would range over real-valued intervals (\eg charge current applied to a battery).
We thus (i)~discretize the controllable values into finitely many control \emph{actions}.
Since (ii)~control decisions are only made every few minutes, the model is discrete-time.
Further, (iii)~wind conditions at the current time are known, so there is a single initial state for every (iv)~finite horizon.
The combination of (i)-(iv) entails that we can only reach a finite subset of the continuous state space.
Thus we can build an MDP with finitely many states and actions.
We use the original (non-linearized) continuous dynamics of the power grid to compute the successor state following an action.
A cost function penalizes frequency deviations;
violations of hard constraints lead to absorbing non-goal states.
An action selection strategy that minimizes the expected accumulated cost to reach the time horizon then defines an optimal deployment of ancillary services.
We track the time horizon, making the MDP a directed acyclic graph in theory and a tree in practice.
In the MPC loop, we only need the action selected in the initial (current) state; when time has advanced to the next control decision, we have a new initial state (based on measured wind conditions in reality and sampled from the wind error DTMC in our experiments) from which to repeat the procedure.
In particular, a large part of the new iteration's MDP had already been computed in the previous iteration; we only need to add one more layer for the advanced time horizon.
We present the continuous-state dynamics of the power system in Sect.~\ref{sec:Grid_model}, explain the formal and technical details of our MDP-based approach in Sect.~\ref{sec:MDP}, and report on an experimental evaluation in Sect.~\ref{sec:Numerical_study}.

Our approach has three key advantages:
We (1) obtain a \emph{strategy that is sound} \wrt the physical constraints, \ie it is guaranteed to satisfy the battery, generator, and ramping capacities (but not necessarily the frequency and transmission limits).
The same cannot be guaranteed with scenario optimization due to the linearization and statistical error.
We (2) \emph{exploit the tree structure} of the MDP to speed up computations in the MPC loop;
and (3) by relying on existing probabilistic model checking technology, the approach is \emph{easy to extend}, \eg with multiple objectives, unreliable communication, or demand uncertainty.
Its main drawbacks are that, while sound and optimal for the discrete MDP, the computed strategy is sound but \emph{may not be optimal for the continuous model}:
an optimal strategy in the continuous model may require a control input that lies between the discrete options of the MDP.
Moreover, the MDP's \emph{state space grows exponentially} with the time horizon and precision of the discretization.
We investigate the effects of varying degrees of discretization and time horizons on the quality of the schedule and the tractability of the problem in our experimental evaluation.

\paragraph{Related work.}
Previous studies of demand-side flexibility consider \eg \emph{vehicle-to-grid}~\cite{Kempton2008,Wang2011} and \emph{buildings-to-grid} integration~\cite{DBLP:journals/tsg/TahaGDPL19,Lymperopoulos2015,DBLP:conf/cdc/RostampourBS19,DBLP:journals/tsg/RazmaraBSPR18}.
As renewable generation is primarily decentralized, regional congestion is an issue~\cite{Pillay2015,Bertsch2016}; congestion management under uncertain generation was studied in~\cite{DBLP:journals/tase/NguyenSB17,Gerard2018,Hemmati2017}.
The majority of the previously cited works use continuous-state models.
Several also apply the MPC pattern~\cite{DBLP:journals/tsg/TahaGDPL19,Liu2018,RostampourBadings2020} and generally state the optimization as an \emph{optimal power flow} problem~\cite{Machowski2006PowerControl}.
As mentioned, scenario optimization~\cite{DBLP:journals/siamjo/CampiG08,DBLP:journals/tac/CalafioreC06,DBLP:journals/tac/MargellosGL14} is sampling-based; in a different approach to sampling,~\cite{DBLP:conf/ccece/Al-SaffarM19} uses Monte Carlo tree search for optimal power flow in the presence of many distributed energy~resources.

When it comes to Markov models,~\cite{grillo2016optimal} uses MDPs for optimal storage scheduling, while~\cite{Tao2020} computes MDP-based optimal charging strategies for electric vehicles.
Probabilistic safety guarantees have been formally verified on DTMC models~\cite{DBLP:conf/qest/PeruffoGPA19,DBLP:journals/tcst/SoudjaniA15}, \ie MDPs where a fixed control strategy is embedded in the model.
\cite{HHB12} studies decentralized protocols in solar panels to stabilize grid frequency.
Finally, we mention that our way of deriving the MDP is similar to the approach of the StocHy tool~\cite{CA19} and its predecessor FAUST$^2$~\cite{SGA15}, which however lack support for costs/rewards and do not implement our efficient MPC loop.

%% file: 0_preliminaries.tex
\section{Preliminaries}\label{sec:preliminaries}
A \emph{discrete probability distribution} over a finite set $X$ is a function $\mu \colon X \to [0,1]$ with $\sum_{x \in X} \mu(x) = 1$.
The set of all distributions over $X$ is $\distr{X}$.
We write $|X|$ for the number of elements in $X$.
Notation $x_{1:n}$ introduces a vector $[x_1, \ldots, x_n]$.

\begin{definition}
    A \emph{Markov decision process} (MDP) is a tuple $\mdp=\MDP$ where $\States$ is a finite set of states, $\Actions$ is a finite set of actions, $\initState \in \States$ is the initial state, $\transfunc \colon \States \times \Actions \rightharpoonup \distr{\States}$ is the (partial) probabilistic transition function, and $\costfunc \colon \States \to \R$ is the state-based cost function.
    We assume deadlock-free MDP.
\end{definition}
A \emph{discrete-time Markov chain} (DTMC) is an MDP with only one action at every state. 
For DTMCs, we omit the set of actions $\Actions$ by simply typing the transition function $\transfunc \colon S \to \distr{S}$.
To define an expected cost measure on MDPs, the nondeterministic choices of actions are resolved by \emph{strategies}.
A memoryless deterministic \emph{strategy} for an MDP is a function $\strat \colon \States \to \Actions$.
For other types of strategies, we refer to~\cite{BK08}.
Applying strategy $\strat$ to an MDP $\mdp$ resolves all nondeterministic choices and yields an \emph{induced DTMC} $\mdp^\strat$.
The expected cost of reaching a set of goal states $G\subseteq \States$ in this induced DTMC is denoted by $\expCost{\mdp^\strat}{G}$.
The goal is to compute a strategy that minimizes the expected cost. 


%% file: 2_power_system_model.tex
\section{Continuous-State Power System Modelling}
\label{sec:Grid_model}

The continuous power system model is a system of nonlinear differential equations.
We explain its setup and components in this section.
We then discretize the model \wrt time, obtaining continuous-state dynamics as a set of nonlinear equations.

\subsection{Grid Model Dynamics}

\begin{figure}[t!]
	\centering
	\scalebox{0.9}{
	\input{Figures/Tikz/Example_grid}
	}
	\caption{3-node example electricity grid.}
	\label{fig:ModellingExampleGrid}
\end{figure}
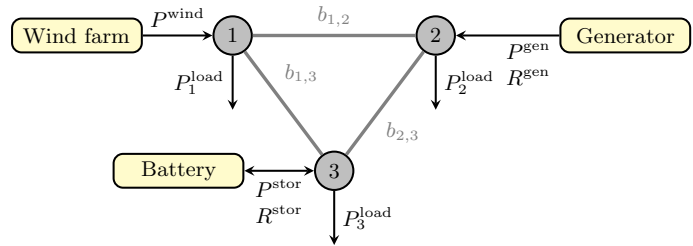

We model a power grid as an undirected graph of interconnected nodes, to which generators and loads are connected.
The example grid in Fig.~\ref{fig:ModellingExampleGrid} has one generator, three nodes with connected loads, plus one wind farm and one battery.
We adopt the grid model dynamics proposed in~\cite{Badings2019,RostampourBadings2020}.
The dynamics in every node are given by the \emph{active power swing equation}, which describes the balance between electrical and kinetic energy at that node in the grid~\cite{DBLP:conf/cdc/TripBP14}.
The state at time $t$ for node $n$ is determined by the voltage angle and frequency, yielding the dynamics:
\begin{equation}
    m_n \ddot{\delta}_n(t) + d_n\dot{\delta}_n(t) = \bar{P}_n(t) - \textstyle{\sum_{p \in \mathcal{N}}}\, b_{n,p}\sin(\delta_n(t) - \delta_p(t))
    \label{eq:SwingEquation}
\end{equation}
where $\delta_n(t)$, $\dot{\delta}_n(t)$, $\ddot{\delta}_n(t) \in \mathbb{R}$ are the voltage angle, angular velocity (frequency), and angular acceleration of node $n$, and $m_n$ and $d_n$ are inertia and damping coefficients, respectively.
$\bar{P}_n(t) \in \mathbb{R}$ is the power balance at node $n$, and is given by the sum of the generation and loads at that node.
The power flow between node $n$ and all connected nodes $p \in \mathcal{N}$ is assumed to be purely reactive and characterized by the line susceptance, $b_{n,p}$, and the difference in voltage angle between the connected nodes.
A detailed description is available in~\cite{Badings2019buildings,RostampourBadings2020}.

\subsection{Grid Frequency Control}

The \textit{grid frequency deviation} for node $n$, denoted $\omega_n(t)$, is the difference between the absolute frequency, $\dot{\delta}_n(t)$ in Eq.~\ref{eq:SwingEquation}, and the desired frequency (\eg $50\,$Hz in Europe).
The value of $\omega_n(t)$ can be controlled by the injection or consumption of electrical energy;
any mismatch between power supply and demand results in a deviation.
As shown in Fig.~\ref{fig:ModellingExampleGrid}, we distinguish five generating or consuming assets: 
\begin{itemize}
\item
$P^\text{gen}(t)$ is the \emph{conventional power dispatch}.
Conventional generators are subject to ramping limits, so the derivative $\dot{P}^\text{gen}(t)$ is restricted to certain bounds.

\item
$P^\text{load}(t)$, the \emph{consumer load}, represents the known and uncontrollable demand.
We can readily extend the model to controllable or uncertain demand.

\item
$R^\text{gen}(t)$ is the deployment of \emph{spinning reserves}.
It is a control variable in the optimal power flow problem.

\item
$P^\text{wind}(t)$, the \emph{wind power generation}, is a random variable~\cite{Papaefthymiou2008} due to its limited predictability.
We define $P^\text{wind,fc}(t)$ as the forecast wind power at time~$t$; then the \textit{forecast error} is $\Delta P^\text{wind}(t) = P^\text{wind}(t) - P^\text{wind,fc}(t)$.

\item
A \emph{battery} is an energy storage buffer whose \textit{state of charge} (SoC) $q(t)$ follows the injection or consumption of energy.
$P_\text{stor}(t)$ is the uncontrollable power input variable, known from the day-ahead plan, and $R_\text{stor}(t)$ is the demand-side flexibility power rate, which is a control variable in the power flow problem.

\end{itemize}

\subsection{Ancillary Service Deployment}
Ancillary service deployment is subject to two restrictions:
reserves and storage flexibility must be scheduled a day ahead,
and the deployment can never exceed the scheduled amount.
The following constraints ensure these restrictions:
\begin{equation}
    -R^\text{gen}_\text{ds}(t) \leq R^\text{gen}(t) \leq R^\text{gen}_\text{us}(t)
    \label{eq:reserve_scheduling}
\end{equation}
ensures that the deployment of spinning reserves is within the scheduled bounds where $R^\text{gen}_\text{ds}(t) \geq 0$ ($R^\text{gen}_\text{us}(t) \geq 0$) is the scheduled amount of down-spinning (up-spinning) reserves. Similarly we have for flexibility that
\begin{equation}
    -R^\text{stor}_\text{dd}(t) \leq R^\text{stor}(t) \leq R^\text{stor}_\text{id}(t),
    \label{eq:flexibility_scheduling}
\end{equation}
where $R^\text{stor}_\text{dd}(k) \geq 0$ and $R^\text{stor}_\text{id}(k) \geq 0$ are the scheduled decreased- and increased-demand flexibility, respectively.

\subsection{Discrete-Time Storage-Integrated Power System Model}
\label{sec:discrete_power_system_model}

We discretize the continuous dynamics with respect to time to render the problem of optimal frequency control tractable.
The resulting model is a set of nonlinear equations, which albeit discretized \wrt time are still defined on continuous state and control spaces.
They describe the transition from one continuous state and control input to the resulting continuous state one discrete time step later.

Consider a power grid with $n_t$ nodes, $n_g$ generators, $n_f$ wind farms, and $n_s$ batteries for storage.
Its continuous dynamics are given as a system of $2n_t + n_g + n_s$ first-order differential equations.
The features of the continuous state space are given by the voltage angles $\delta_{1:n_t}$ and frequencies $\omega_{1:n_t}$ for all $n_t$ nodes, the power generation $P^\text{gen}_{1:n_g}$ for all $n_g$ generators, and the state of charge $q_{1:n_s}$ of all $n_s$ batteries.
The vector of control variables contains the \emph{change} in generator dispatch $\dot{P}^\text{gen}_{1:n_g}$ and the reserve deployment $R^\text{gen}_{1:n_g}$ for all $n_g$ generators, plus the flexibility deployment $R^\text{stor}_{1:n_s}$ for all $n_s$ batteries.
We discretize \wrt time via the first-order backward Euler implicit method~\cite{Sincovec1981} to obtain the nonlinear function
\begin{equation}
    x(k+1) = f \big( x(k), u(k), v(k), w(k) \big),
    \label{eq:full_discrete_statespace}
\end{equation}
\begin{itemize}
    \item with $f(\cdot)$ reflecting the dynamics of the considered power system, which are nonlinear due to the sinusoid, $\sin(\delta_n(t) - \delta_p(t))$, in Eq.~\ref{eq:SwingEquation},
    \item $x(k) = [\delta_{1:n_t}, \omega_{1:n_t}, P^\text{gen}_{1:n_g}, q_{1:n_s}] \in \mathbb{R}^{2n_t + n_g + n_s}$ the state vector,
    \item $u(k) = [\dot{P}^\text{gen}_{1:n_g}, R^\text{gen}_{1:n_g}, R^\text{stor}_{1:n_s}] \in \mathbb{R}^{2n_g + n_s}$ the vector of control variables,
    \item $v(k) = [P^\text{load}_{1:n_t}\!, P^\text{wind,fc}_{1:n_f}\!, P^\text{stor}_{1:n_s}] \,{\in}\, \mathbb{R}^{n_t + n_f + n_s}$ the uncontrollable known inputs,~and
    \item $w(k) = [\Delta P^\text{wind}_{1:n_f}] \in \mathbb{R}^{n_f}$ the vector of uncontrollable random variables.
\end{itemize}
We omit further details for the sake of brevity and refer the interested reader to~\cite{Badings2019} and~\cite{DBLP:journals/tsg/TahaGDPL19} for the full derivation and discretization of similar grid models.

\subsubsection{Power balance and control variables.}
\label{sec:Grid_model_constraints}

The day-ahead generator power dispatch is scheduled such that generation plus wind power forecast matches the consumer load pattern.
We impose the following constraint at every time point $k$:
\begin{equation}
    \textstyle\sum_{i = 1}^{n_g} P^\text{gen}_i (k)
    + \textstyle\sum_{m = 1}^{n_f} P^\text{wind,fc}_m (k)
    = \textstyle\sum_{n = 1}^{n_t} P^\text{load}_n (k).
    \label{eq:Conventional_balance}
\end{equation}
Since $P^\text{wind,fc}_m$ and $P^\text{load}_n$ are known, imposing this equality constraint yields $n_g - 1$ independent control variables.
Hence, in a single-generator grid, the day-ahead planning is fixed by wind forecast and consumer load, while in a grid with multiple generators, the required total dispatch must be divided between the different units.
In a similar manner, during the day itself, the reserve power and storage flexibility can be deployed together to restore the mismatch in the power balance caused by forecast errors at any time point:
\begin{equation}
    \textstyle\sum_{i = 1}^{n_g} R^\text{gen}_i (k)
    + \textstyle\sum_{m = 1}^{n_f} \Delta P^\text{wind}_m (k)
    = \textstyle\sum_{n = 1}^{n_s} R^\text{stor}_n (k),
	\label{eq:ReservePower_balance}
\end{equation}
where $\Delta P^\text{wind}_m$ is a random variable, and both $R^\text{gen}_i$ and $R^\text{stor}_n$ are control variables.
Hence, imposing Eq.~\ref{eq:ReservePower_balance} yields $n_g + n_s - 1$ independent control variables.

\subsubsection{Power system constraints.}
The discrete-time power system model in Eq.~\ref{eq:full_discrete_statespace} is subject to a number of constraints.
First of all, the equality constraints in Eqs.~\ref{eq:Conventional_balance} and \ref{eq:ReservePower_balance} are imposed to enforce the balance between power supply and demand.
Second, power lines have limited transmission capacity, and generators have limited generation capacity and ramping capability.
Third, the deployment of reserve power and storage flexibility is limited by their scheduled values, as described in Eqs.~\ref{eq:reserve_scheduling} and \ref{eq:flexibility_scheduling}.
Finally, the electrical storage units have a limited capacity, and can only be charged or discharged at a given maximum rate.
For the explicit formulation of the constraints we refer the interested reader to~\cite{RostampourBadings2020}.

%% file: Figures/Tikz/Example_grid.tex
\tikzstyle{bus} = [circle, minimum width=0.5cm, text centered, draw=black, fill=gray!50]

\tikzstyle{smallNode} = [rectangle, rounded corners, minimum width=1.7cm, text width=1.7cm, minimum height=0.5cm, text centered, draw=black, fill=yellow!25]


\begin{tikzpicture}[node distance=2cm,>=stealth,line width=0.3mm,auto,
	main node/.style={circle,draw,font=\sffamily\Large\bfseries}]
	
	\newcommand\xshift{1.1cm}
	
	\node (bus3) [bus] {$3$};
	\node (bus1) [bus, above of=bus3, xshift=-1.5cm] {$1$};
	\node (bus2) [bus, above of=bus3, xshift=1.5cm] {$2$};
	
	\node (generator2) [smallNode, right of=bus2, xshift=0.8cm] {Generator};
	\node (windfarm1) [smallNode, left of=bus1, xshift=-0.3cm] {Wind farm};
	
	\node (storage3) [smallNode, left of=bus3, xshift=-0.3cm] {Battery};
	
	\draw[-, gray, line width=0.5mm] (bus1) -- (bus2) node[pos=0.5, align=center] {$b_{1,2}$};
	\draw[-, gray, line width=0.5mm] (bus1) -- (bus3) node[pos=0.4, align=center] {$b_{1,3}$};
	\draw[-, gray, line width=0.5mm] (bus2) -- (bus3) node[pos=0.6, align=center] {$b_{2,3}$};
	
	\draw[->] (bus1.south) -- ($(bus1.south) + (0cm,-0.8cm)$) node[pos=0.5, align=center, anchor=east] {$P^{\text{load}}_1$};
	\draw[->] (bus2.south) -- ($(bus2.south) + (0cm,-0.8cm)$) node[pos=0.5, align=center] {$P^{\text{load}}_2$};
	\draw[->] (bus3.south) -- ($(bus3.south) + (0cm,-0.8cm)$) node[pos=0.5, align=center] {$P^{\text{load}}_3$};
	
	\draw[<->] (bus3.west) -- (storage3.east) node[pos=0.5, align=center] {$P^{\text{stor}}$ \\ $R^{\text{stor}}$};
	
	\draw[->] (windfarm1.east) -- (bus1.west) node[pos=0.5, align=center] {$P^{\text{wind}}$};
	\draw[->] (generator2.west) -- (bus2.east) node[pos=0.3, align=center] {$P^{\text{gen}}$ \\ $R^{\text{gen}}$};
	
	
	
	
    
\end{tikzpicture}

%% file: 3_discrete_MDP.tex
\section{Discrete-State Receding Horizon Control Problem}
\label{sec:MDP}

Our goal is to overcome the need for sampling and linearization to optimize ancillary service deployment.
Directly using the model presented in Eq.~\ref{eq:full_discrete_statespace} would require dealing with continuous state and action spaces.
Our approach is to discretize the actions, then explore the resulting finite number of (continuously-valued) successor states up to a given exploration depth.
In this way, we obtain a finite MDP that can be solved iteratively using a receding horizon principle.
This approach is a discrete-state model predictive control technique.
We now present the wind error DTMC, followed by the details of our exploration procedure and the formal definition of the finite-horizon MDP.

\subsection{Stochastic Wind Power Model}\label{sec:wind_model}
\label{sec:MDP_DTMC}

Recall that the wind power forecast error, $\Delta P^\text{wind}_{m}(k) \in \mathbb{R}$, of every individual wind farm $m \in \{ 1, \ldots, n_f \}$ is a continuous random variable in Eq.~\ref{eq:full_discrete_statespace}.
Exploiting the time discretization of Sect.~\ref{sec:discrete_power_system_model}, we construct a DTMC for the forecast error of every wind farm with time resolution equal to the discretization level of the dynamics.
For this section, we assume the presence of only one wind farm to simplify notation.
Let $S$ be the finite state space of the DTMC, and let function $M\colon \mathbb{R} \to S$ map the wind power forecast error $\Delta P^\text{wind} \in \mathbb{R}$ to a state $s_w \in S$:
every value of the continuous random variable $\Delta P^\text{wind}$ is approximated by the value of a discrete state $s_w \in \States$.
The DTMC's transition function $\transfunc \colon S \to \distr{S}$ describes the probability that a transition occurs from one state $s_w$ at any time $k$ to another state $s_w'$ at time $k+1$ with probability $\transfunc(s_w)(s_w')$.

\subsubsection{Historical wind power data.}

\begin{figure*}[t!]
	\begin{subfigure}{0.50\textwidth}
		\includegraphics[width=\linewidth]{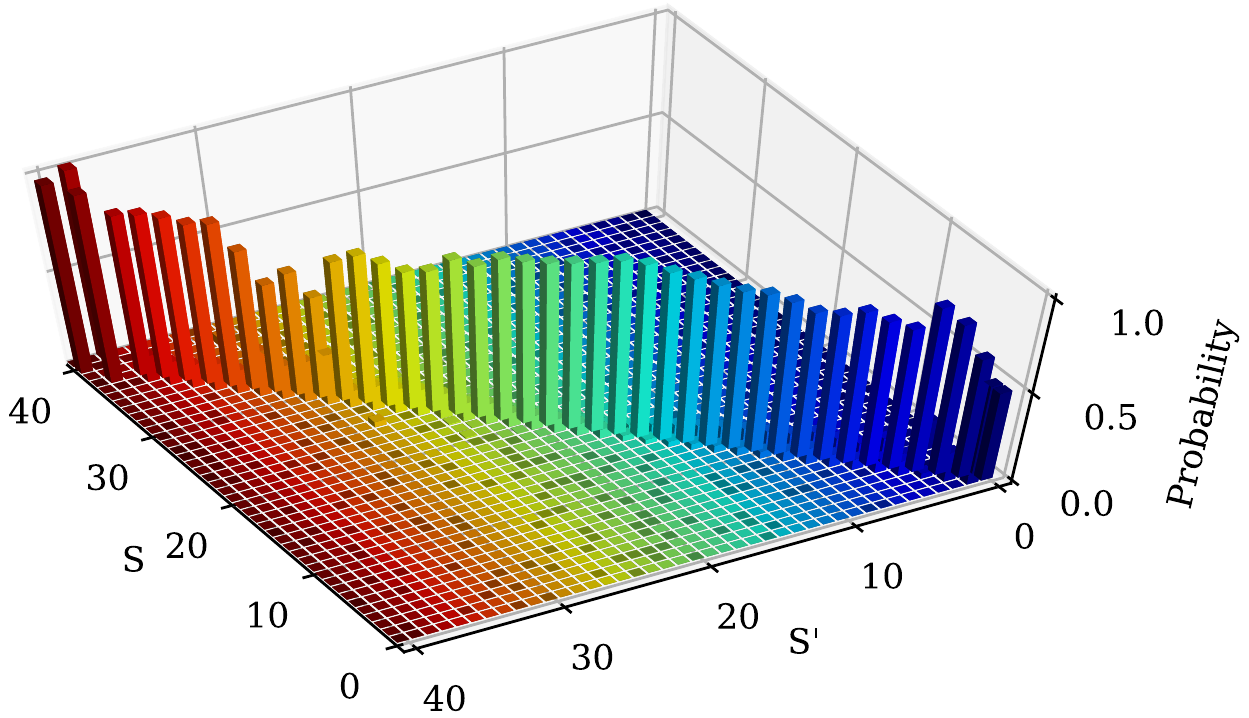}
		\caption{Using 5-minute based data.}
		\label{fig:wind_transition_matrix300}
	\end{subfigure}
	\begin{subfigure}{0.50\textwidth}
		\includegraphics[width=\textwidth]{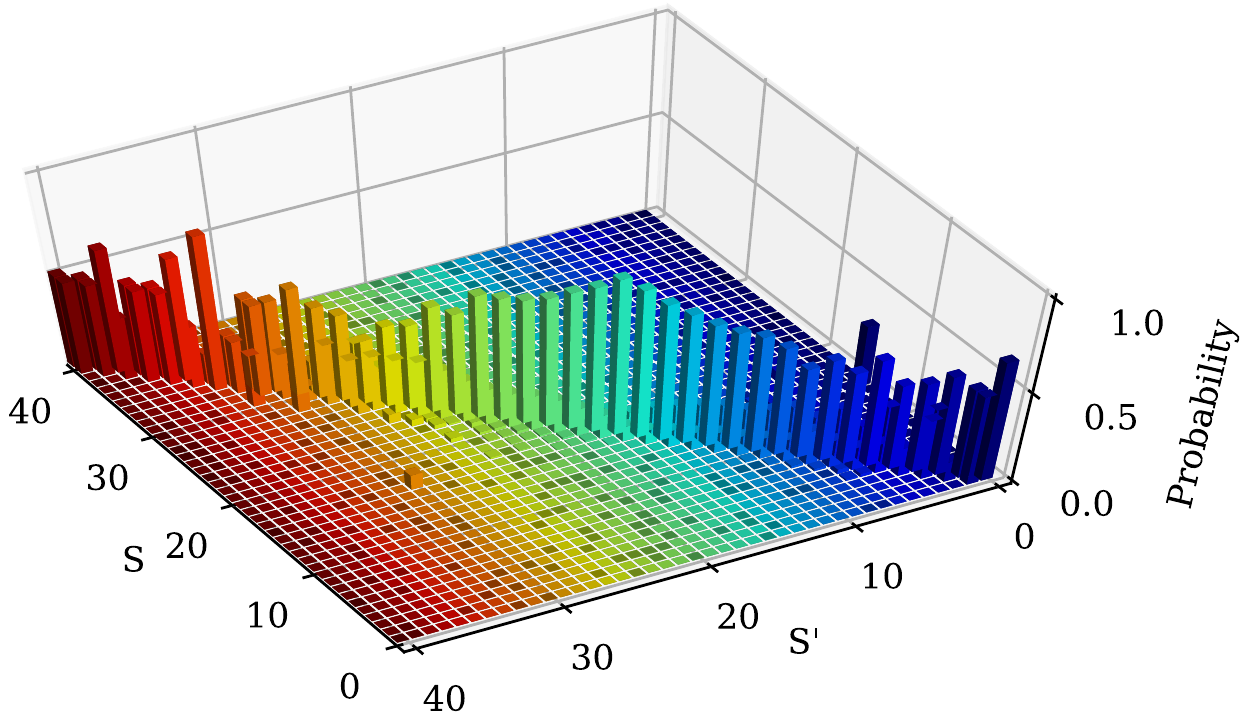}
		\caption{Using 15-minute based data.}
		\label{fig:wind_transition_matrix900}
	\end{subfigure}
	
	\caption{Transition probability function of the wind error DTMC.}
	\label{fig:wind_transition_matrix}
\end{figure*}

We follow the method proposed in~\cite{Papaefthymiou2008,Margellos2012} to construct the DTMC based on the historical wind power forecast error.
The states $\States$ are based on a uniform discretization of the historical forecast error. 
The transition probabilities in $\transfunc$ are determined using a maximum likelihood estimation, by counting the number of transitions from one state $s_w$ at time $k$ to a successor state $s_w'$ at time $k+1$.
We use five years (2015-2019) of on-shore wind power data measured every 15 minutes from the TenneT region of the German transmission grid, obtained from the \emph{ENTSO-E Transparency Platform}~\cite{ENTSOe}.
We interpolate the data to a 5-minute basis, to match the time resolution we employ in the numerical demonstration in Sect.~\ref{sec:Numerical_study}.
The data set contains the wind power forecast, $P^\text{wind,fc}$, and the actual power, $P^\text{wind}$, thus providing a five-year time series of the observed wind power error, $\Delta P^\text{wind}$.
$M$ then maps every continuous value of $\Delta P^\text{wind}$ to one of $41$ discrete states, \ie $S = \{ s^1_{w}, \ldots, s^{41}_{w} \}$, as in~\cite{Margellos2012}.
We show the resulting transition matrix in Fig.~\ref{fig:wind_transition_matrix} for the original 15-minute and the interpolated 5-minute data.
The matrix is diagonally dominant, reflecting the strong auto-correlation of the forecast error.

\subsection{State Space Exploration Procedure}

Next, we present our method to explore the continuous state space of the model up to a given exploration horizon.
We discretize the continuous control variables such that their potential values define a set of actions for an MDP, then a priori eliminate those actions that lead to states violating the constraints of Sect.~\ref{sec:Grid_model_constraints}.

Let $\mathsf{X}_k = (x(k), s_w)$ denote the continuous state vector $x(k)$ at time $k$, and the DTMC state $s_w$ associated with the current wind power forecast error.
To define an initial state, we use a concrete measurement at the first time step. 
Then, we use the dynamics in Eq.~\ref{eq:full_discrete_statespace} in combination with transitions of the wind error DTMC to compute the possible successor states for each time step.
At time step $k$ we select a control input $u(k)$ and, based on the dynamics in Eq.~\ref{eq:full_discrete_statespace}, receive a set of successor states with different $x(k+1)^1, \ldots, x(k+1)^n$, one associated with every possible wind error successor state, $s^1_w,\ldots,s^n_w$ with $T(s_w)(s^i_w)>0$ for all $1\leq i\leq n$.
The features in $x(k+1)^i$ related to power generation $(P^\text{gen}_{1:n_g})$ and battery SoC $(q_{1:n_s})$ are equal for all $1\leq i\leq n$, while the grid features $(\delta_{1:n_t}, \omega_{1:n_t})$ depend on the wind successor state, $s^i_w$.
As the probabilities to reach these successor states depend exclusively on the probabilities defined by the DTMC, we reach state $\mathsf{X}_{k+1} = (x(k+1)^i, s^i_w))$ with probability $T(s_w)(s^i_w)$.

\subsubsection{Feasible control space.}
Under the two balance constraints in Eqs.~\ref{eq:Conventional_balance} and \ref{eq:ReservePower_balance}, the vector of control variables $u(k)$ in Eq.~\ref{eq:full_discrete_statespace} contains $2 n_g + n_s - 2$ independent variables.
Since the dynamics in Eq.~\ref{eq:full_discrete_statespace} and the constraints imposed on the system are known, we can determine the continuous subset of control inputs $u \in \mathcal{U}_{\mathsf{X}_k} \subset \mathbb{R}^{2 n_g + n_s -2}$ that do not lead to a violation of any of the constraints at time $k+1$.
Note that this set depends on the state $\mathsf{X}_k$ at time $k$.
Given the subset of feasible continuous control inputs, we apply a grid-based discretization in all $2n_g + n_s - 2$ dimensions, to obtain a set $A_{\mathsf{X}_k}$ of \emph{feasible and discrete actions} at time $k$, where the subscript denotes the dependency on $\mathsf{X}_k$.
From this discretization of the control actions, an important property follows.
Given the current state $\mathsf{X}_k$ at time $k$, every action $a \in A_{\mathsf{X}_k}$ has a different set of \emph{discrete successor states} $\mathsf{X}_{k+1}$ at time $k+1$.
By only exploring the continuous state-space for the feasible actions in $A_{\mathsf{X}_k}$, we minimize the size of the resulting finite-state model.

\begin{figure}[t!]
	\centering
	\scalebox{0.85}{
	\input{Figures/Tikz/SchematicActions}
	}
	\caption{Discrete actions $a_1, \ldots, a_5$ and their mapping to two continuous state-space features: a battery SoC and the frequency in one node. Curved arrows show the effect of wind successor states, and dashed lines are system constraints.}
	\label{fig:schematicActons}
\end{figure}
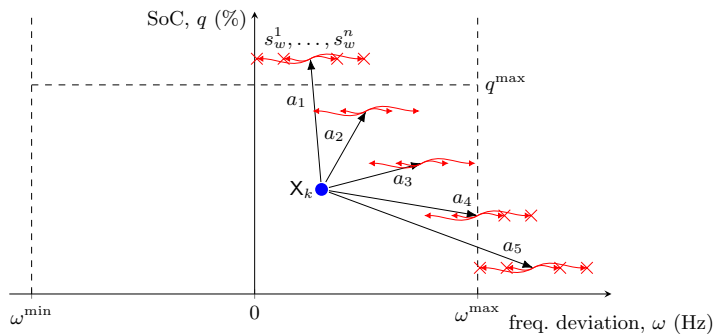

A schematic example of this procedure for two state features (one node frequency and one battery SoC) is shown in Fig.~\ref{fig:schematicActons}. 
The blue dot corresponds to current state $\mathsf{X}_k$ at time $k$, and the straight arrows show the discrete actions $a \in A_{\mathsf{X}_k}$.
The curved arrows show the effect of the forecast error to the grid frequency, which depends directly on the actual successor state in the DTMC.
Because all successor states of $a_1$ violate the maximum SoC constraint ($q(k) \leq q^\text{max}$), this action can be \emph{eliminated a priori}.
Similarly, all successors for $a_5$ violate the maximum frequency deviation limit ($\omega(k) \leq \omega^\text{max}$).
Actions $a_2$ and $a_3$ will not lead to any violation, and are, therefore, included as feasible discrete actions.
Action $a_4$ may or may not violate the constraints depending on the wind power forecast error.
Therefore, this action cannot be eliminated a priori.

\subsubsection{Exploration.}
\begin{figure}[t!]
	\centering
	\input{Figures/Tikz/ExplorationTree}

	\caption{Visualization of the state-space exploration procedure, with initial measurement $\mathsf{X}_k$ at time $k$, and finite horizon with end time $k+K_h$. Solid edges represent discrete actions, whereas dashed lines reflect different successor states.}
	\label{fig:ExplorationTree}
\end{figure}
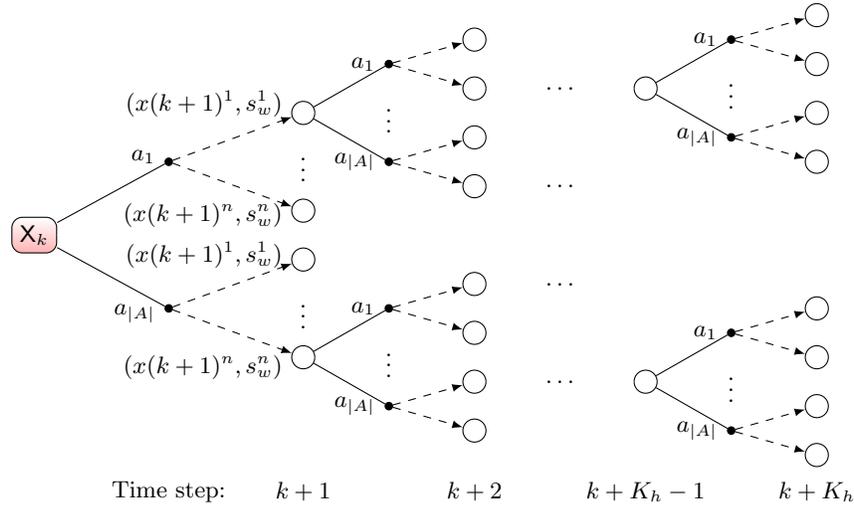

The finite exploration horizon with starting time $k$ and look-ahead of $K_h$ steps is the set $\{ k, \ldots, k+K_h \}$.
The exploration procedure can be performed recursively for all discrete actions $a \in A_{\mathsf{X}_k}$ for every $k$, until reaching the desired horizon.
We obtain a tree-structured model as in Fig.~\ref{fig:ExplorationTree}, where the \emph{branching factor} depends on the number of actions and possible wind successor states, and the \emph{depth} is given by the horizon length.
For brevity, the dependency of the set of discrete actions on the current state is omitted in this figure.

Recall that we consider only those states $\mathsf{X}_k = (x(k), s_w)$ of the continuous state space that are visited during the exploration.
We provide the full definition of the MDP $\MDP$ with exploration horizon $\{ k, \ldots k+K_h \}$, where
\begin{itemize}
    \item every state $s \in \States$ is associated with an $\mathsf{X}_k$ from the continuous state space;
    \item the set of actions $A$ is the union of all feasible action sets $A_{\mathsf{X}_k}$ for all $\mathsf{X}_k$;
    \item the initial state $s_I$ is given by a concrete measurement $(x_0, s_{w})$;
    \item the partial probabilistic transition function $\transfunc \colon \States \times \Actions \to \distr{\States}$ maps every state and action to the corresponding distribution over successor states according to the wind error DTMC;
    \item the cost function $\costfunc \colon \States \to \R$ assigns the immediate cost given by the sum of the absolute value of the frequency deviation in every grid node.
\end{itemize}
Finally, we define the set $G\subseteq S$ of \emph{goal states} as the states that are reached at the end of the horizon $k+K_h$ and satisfy the constraints.

\subsubsection{Receding horizon and tree structure.}

We compute the strategy $\strat$ that induces the minimal expected cost $\expCost{\mdp^\strat}{G}$ for the MDP with horizon $\{ k, \ldots k+K_h \}$.  
Using this strategy, we implement the (first) optimal action $a_0=\strat(s_0)$ where $s$ is associated with $\mathsf{X}_k$. 
Having executed the optimal action at time $k$, we then apply the so-called \emph{receding horizon principle}, meaning that we \emph{shift} the exploration horizon one step forward in time.
We update the MDP for the shifted horizon, which is now given as $\{ k+1, \ldots k+K_h+1 \}$, and again compute an optimal strategy.
Hence, with every shift of the exploration horizon in time, we \emph{reveal small bits of new information} near the end of the horizon.

As an example, consider an initial exploration horizon defined as the time frame between 9:00-9:15 AM.
Computing an optimal strategy for the corresponding MDP means that we take all the information within that time frame into account, \ie we have perfect knowledge within the horizon.
However, a possible change in the power demand (or any other uncontrollable input) that is forecast at 9:30 AM is not revealed to the model, until the exploration horizon also spans that time step.
Defining an adequate length of the exploration horizon reflects a trade-off between model size and the optimality of the solution.

When we shift the exploration horizon from $\{ k, \dots, k+K_h \}$ to $\{ k+1, \dots, k+K_h+1 \}$, another layer is added to the MDP. 
As the starting time of the horizon also shifts, we gain new information via a new measurement at $k+1$, and then obtain a single new initial state for the MDP with the new horizon. 
By exploiting the tree structure of the MDP, we extend the current MDP with a new layer of successor states and take the subtree starting at the newly found initial state at time $k+1$.
Simultaneously, we shrink the layer at the top, by pruning all states that cannot be reached anymore from the new initial state.

The approach of iteratively solving and updating the MDP for shifting horizons is similar to a discrete-state model predictive control technique.
This method is applied iteratively, until a desired simulation horizon is reached (\eg 24 hours).

%% file: Figures/Tikz/SchematicActions.tex
\usetikzlibrary{shapes.geometric, arrows}
\usetikzlibrary{arrows.meta, shapes, positioning, calc}

\tikzset{cross/.style={cross out, draw=red, minimum size=2*(#1-\pgflinewidth), inner sep=0pt, outer sep=0pt},
cross/.default={3pt}}

\begin{tikzpicture}
\begin{axis}[
    axis y line=middle, 
    axis x line=bottom,
    xlabel = {freq. deviation, $\omega$ (Hz)},
    ylabel = { SoC, $q$ (\%)},
    every axis x label/.style={
    at={(ticklabel* cs:1)},
    anchor=north,
    yshift=-0.2cm
    },
    every axis y label/.style={
    at={(ticklabel* cs:0.97)},
    anchor=east,
    },
    width=11cm,
    height=6cm,
    xtick = {-1,0,1},
    xticklabels = {$\omega^\text{min}$, $0$, $\omega^\text{max}$},
    ytick = {0},
    xmin=-1.1,
    xmax=1.6,
    ymin=0,
    ymax=1.08,
]

\draw [dashed] ({axis cs:1,0}|-{rel axis cs:0,0}) -- ({axis cs:1,0}|-{rel axis cs:0,1});
\draw [dashed] ({axis cs:-1,0}|-{rel axis cs:0,0}) -- ({axis cs:-1,0}|-{rel axis cs:0,1});
\draw [dashed] ({axis cs:-1,0.8}) -- ({axis cs:1,0.8}) node[pos=1, anchor=west] {$q^\text{max}$};

\node (init) at ({axis cs:0.3,0.4}) [] {};
\node at (init) [anchor=east] {$\mathsf{X}_k$};
\fill [blue] ({axis cs:0.3,0.4}) circle (0.1cm);

\draw ({axis cs:0.25+0.24, 0.9}) node[cross] {};
\draw ({axis cs:0.25+0.12, 0.9}) node[cross] {};
\draw ({axis cs:0.25-0.12, 0.9}) node[cross] {};
\draw ({axis cs:0.25-0.24, 0.9}) node[cross] {};

\draw ({axis cs:1.25+0.24, 0.1}) node[cross] {};
\draw ({axis cs:1.25+0.12, 0.1}) node[cross] {};
\draw ({axis cs:1.25-0.12, 0.1}) node[cross] {};
\draw ({axis cs:1.25-0.24, 0.1}) node[cross] {};

\draw ({axis cs:1.00+0.24, 0.3}) node[cross] {};
\draw ({axis cs:1.00+0.12, 0.3}) node[cross] {};


\foreach \i in {1,...,2}
{
    \edef\temp{\noexpand\draw[-{Latex}] (init) -- ({axis cs:\i*0.25, -\i*0.2+1.1}) node[auto, pos=0.65, anchor=east] {$a_\i$};}
    \temp
    
    \foreach \j in {-2,-1,1,2}
    {
        \edef\temp{\noexpand\draw[-{Latex[length=1mm,width=1mm]}, red, bend right=90] ({axis cs:\i*0.25, -\i*0.2+1.1}) to [out=30,in=180] ({axis cs:\i*0.25-\j*0.12, -\i*0.2+1.1});}
        \temp
    }
}

\foreach \i in {3}
{
    \edef\temp{\noexpand\draw[-{Latex}] (init) -- ({axis cs:\i*0.25, -\i*0.2+1.1}) node[auto, pos=0.8, anchor=north] {$a_\i$};}
    \temp
    
    \foreach \j in {-2,-1,1,2}
    {
        \edef\temp{\noexpand\draw[-{Latex[length=1mm,width=1mm]}, red, bend right=90] ({axis cs:\i*0.25, -\i*0.2+1.1}) to [out=30,in=180] ({axis cs:\i*0.25-\j*0.12, -\i*0.2+1.1});}
        \temp
    }
}

\foreach \i in {4,...,5}
{
    \edef\temp{\noexpand\draw[-{Latex}] (init) -- ({axis cs:\i*0.25, -\i*0.2+1.1}) node[auto, pos=0.9, anchor=south] {$a_\i$};}
    \temp
    
    \foreach \j in {-2,-1,1,2}
    {
        \edef\temp{\noexpand\draw[-{Latex[length=1mm,width=1mm]}, red, bend right=90] ({axis cs:\i*0.25, -\i*0.2+1.1}) to [out=30,in=180] ({axis cs:\i*0.25-\j*0.12, -\i*0.2+1.1});}
        \temp
    }
}

\node[align=center] at ({axis cs:0.25, 0.98}) {$s^1_w,\ldots,s^n_w$};

\end{axis}
\end{tikzpicture}

%% file: Figures/Tikz/ExplorationTree.tex
\tikzset{
  treenode/.style = {shape=rectangle, rounded corners,
                     draw, align=center,
                     top color=white, bottom color=blue!20},
  root/.style     = {treenode, bottom color=red!30},
  env/.style      = {treenode},
  state/.style    = {circle,draw},
  action/.style   = {circle, draw, inner sep=1pt, fill=black}
}

\begin{tikzpicture}
  [
    grow                    = right,
    level 1/.style={level distance=5.5em, sibling distance=3em}, 
    level 2/.style={level distance=5.5em, sibling distance=2em}, 
    level 3/.style={level distance=3.5em, sibling distance=2em},
    level 4/.style={level distance=3.5em, sibling distance=1.5em}
    level 5/.style={level distance=3.5em, sibling distance=1.5em},
    edge from parent/.style = {draw},
    every node/.style       = {font=\footnotesize},
    sloped
  ]
  \node [root] {$\mathsf{X}_k$}
    child { node [action] {} 
        child [dashed, -latex] { node [state, solid] {} 
            child [solid,-] { node [action] {} 
                child [dashed, -latex] { node [state, solid] {} }
                child [dashed, -latex] { node [state, solid] {} 
                    child[color=white] { node [text=black] {$\ldots$} 
                        child [solid,-] { node [color=black, state] {} 
                            child[color=black] { node [action] {} 
                                child [dashed, -latex] { node [state, solid] {} }
                                child [dashed, -latex] { node [state, solid] {} }
                            }
                            child[color=white] { node [text=black] {$\vdots$} }
                            child[color=black] { node [action] {}
                                child [dashed, -latex] { node [state, solid] {} }
                                child [dashed, -latex] { node [state, solid] {} }
                            }
                        }
                    }
                }
            }
            child[color=white] { node [text=black] {$\vdots$} }
            child [solid,-] { node [action] {} 
                child [dashed, -latex] { node [state, solid] {} }
                child [dashed, -latex] { node [state, solid] {} 
                    child[color=white] { node [text=black] {$\ldots$} }
                }
            }
        }
        child[color=white] { node [text=black] {$\vdots$} }
        child [dashed, -latex] { node [state, solid] {} }
    }
    child[color=white] { node [text=black] {} }
    child { node [action] {} 
        child [dashed, -latex] { node [state, solid] {} }
        child[color=white] { node [text=black] {$\vdots$} }
        child [dashed, -latex] { node [state, solid] {} 
            child [solid,-] { node [action] {} 
                child [dashed, -latex] { node [state, solid] {} 
                    child[color=white] { node [text=black] {$\ldots$} }
                }
                child [dashed, -latex] { node [state, solid] {} }
            }
            child[color=white] { node [text=black] {$\vdots$} }
            child [solid,-] { node [action] {} 
                child [dashed, -latex] { node [state, solid] {} 
                    child[color=white] { node [text=black] {$\ldots$} 
                        child [solid,-] { node [color=black, state] {} 
                            child[color=black] { node [action] {} 
                                child [dashed, -latex] { node [state, solid] {} }
                                child [dashed, -latex] { node [state, solid] {} }
                            }
                            child[color=white] { node [text=black] {$\vdots$} }
                            child[color=black] { node [action] {}
                                child [dashed, -latex] { node [state, solid] {} }
                                child [dashed, -latex] { node [state, solid] {} }
                            }
                        }
                    }
                }
                child [dashed, -latex] { node [state, solid] {} }
            }
        }
    };
    
    \node at (5.3em, 3.2em) (a0) [anchor=east] {$a_1$};
    \node at (5.3em, -3.2em) (a0) [anchor=east] {$a_{|A|}$};
    
    \node at (14.3em, 7em) (a0) [anchor=east] {$a_1$};
    \node at (14.3em, 3.0em) (a0) [anchor=east] {$a_{|A|}$};
    
    \node at (14.3em, -7em) (a0) [anchor=east] {$a_{|A|}$};
    \node at (14.3em, -3.0em) (a0) [anchor=east] {$a_1$};
    
    \node at (28.3em, 8.0em) (a0) [anchor=east] {$a_1$};
    \node at (28.3em, 4.0em) (a0) [anchor=east] {$a_{|A|}$};
    
    \node at (28.3em, -8.0em) (a0) [anchor=east] {$a_{|A|}$};
    \node at (28.3em, -4.0em) (a0) [anchor=east] {$a_1$};
    
    \node at (10.5em, 5.4em) (a0) [anchor=east] {$(x(k+1)^1, s^1_w)$};
    \node at (10.5em, 0.8em) (a0) [anchor=east] {$(x(k+1)^n, s^n_w)$};
    
    \node at (10.5em, -5.4em) (a0) [anchor=east] {$(x(k+1)^n, s^n_w)$};
    \node at (10.5em, -0.8em) (a0) [anchor=east] {$(x(k+1)^1, s^1_w)$};
    
    \node at (5.5em, -10.5em) (a0) [align=center] {Time step:};
    \node at (11.0em, -10.5em) (a0) [align=center] {$k+1$};
    \node at (18.0em, -10.5em) (a0) [align=center] {$k+2$};
    \node at (25.0em, -10.5em) (a0) [align=center] {$k+K_h-1$};
    \node at (32.0em, -10.5em) (a0) [align=center] {$k+K_h$};
    
\end{tikzpicture}

%% file: 4_numerical_study.tex
\section{Numerical Study}
\label{sec:Numerical_study}
We demonstrate the performance of our approach on multiple variants of the $3$-node example grid already introduced in Fig.~\ref{fig:ModellingExampleGrid}.
To this end, we \textit{simulate the wind error DTMC}, and apply our method to build the MDP using the sampled wind power forecast error at every time step.
As such, every wind power trajectory sampled from the DTMC results in a (potentially different) run of our approach.
We solve the MDP to obtain the first action in the sequence of optimal decisions over the exploration horizon, which is then executed to obtain the initial state at the next time step.
By applying the receding horizon principle, we iteratively follow this procedure, until a final simulation horizon of 24 hours is reached.

\subsection{Experimental Setup}

\begin{figure}[t!]
	\centering
	\scalebox{0.82}{\includegraphics{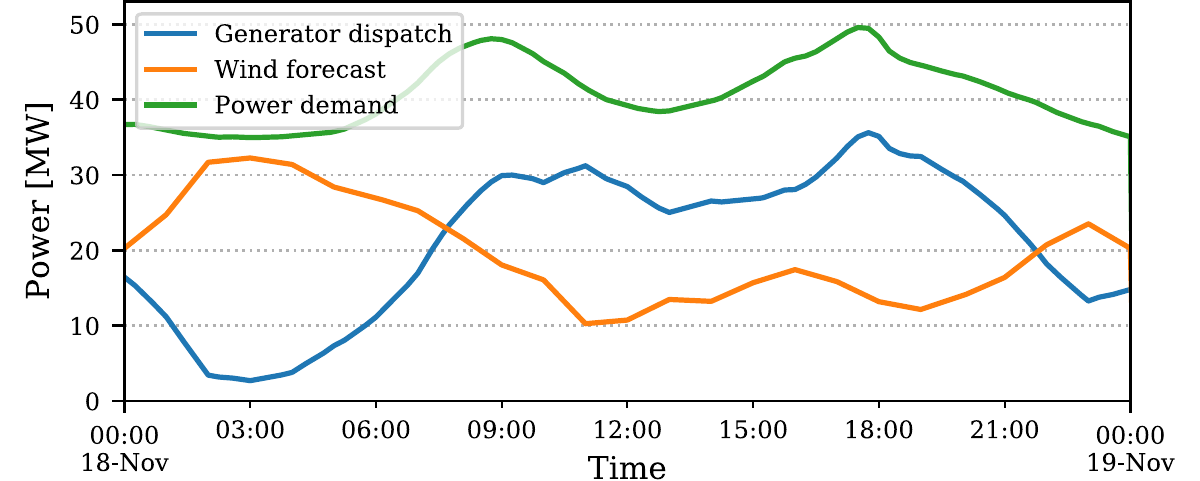}}
	
	\caption{The day-ahead power balance between generator dispatch, wind power forecast, and the power demand, which is equivalent for all performed simulations.}
	\label{fig:Results_day-ahead_balance}
\end{figure}

We consider a simulation time resolution of $5$ minutes, and a full simulation horizon of $24$ hours.
Using the receding horizon exploration procedure, this means that $\frac{60}{5} \cdot 24 = 288$ MDPs are solved to obtain the results over one $24$-hour run.
The same demand and wind power forecast are used for every simulation, resulting in the day-ahead power dispatch shown in Fig.~\ref{fig:Results_day-ahead_balance}.
This figure shows that under the forecast conditions, the power supply and demand are perfectly balanced, leading to a stable grid frequency in the absence of forecast errors.
Both the power demand and wind power forecast are based on historical data obtained from the \emph{ENTSO-E Transparency Platform}~\cite{ENTSOe}, and are scaled appropriately for the simulation study.
The scheduled flexibility deployment limits per battery are set to $\pm 2$ MW, while the scheduled reserve limits per generator are $\pm 0.25$ MW.

\subsubsection{Simulation cases.}

Simulations are performed on three variants of the example grid in Fig.~\ref{fig:ModellingExampleGrid}. In the simplest case, we aggregate the three nodes into one, resulting in a grid where all assets are connected to the same node. The second variant is the exact network shown in Fig.~\ref{fig:ModellingExampleGrid}. The third variant is an extension of the second, where a second storage unit is connected to node 1.

Then, we perform simulation studies for different values of: a) the \emph{exploration horizon} ($300$, $600$, and $900$ seconds), and b) the \emph{levels of discretization for the actions} denoted by $\lambda$ (with values between $3$ and $25$ steps).
To evaluate the quality of the solution of our technique, we perform a statistical analysis on every case, by repeating every experiment $100$ times.
Note that this Monte Carlo type simulation is merely used to evaluate the quality of the obtained solutions, and not required to apply our technique in practice.

\subsubsection{Source code.}
All of the code and input data needed to reproduce the results presented in this paper are available at \url{https://gitlab.science.ru.nl/tbadings/power-system-mdp}.
Our prototype implementation is written in Python version 3.8.3, and allows the user to run the cases defined above or simulate with any other parameter setting.
The simulations are run on a Windows laptop with a 1.30GHz Intel(R) i7-1065G7 CPU and 16.0 GB of RAM.

\subsection{Results}

\subsubsection{Run times.}
The observed run times are approximately proportional to the number of MDP states, and grow exponentially with the exploration horizon.
For the $3$-node system with $1$ battery, an action discretization level of $\lambda=5$ steps, and exploration horizon of $300$ seconds, the average run time \emph{per iteration} of the receding horizon (\ie for solving one MDP) is $0.01$ seconds, resulting in around $3.51$ seconds per $24$h run.
For the same case with longer exploration horizons of $600$ and $900$ seconds, average run times are $0.14$ and $2.09$ seconds, respectively, per receding horizon iteration.

The strongest increase in run time is observed for the $3$-node, $2$ battery case, where the average observed times were $0.03$, $1.03$, and $223.59$ seconds for exploration horizons of $300$, $600$, and $900$ seconds, respectively.
This steeper increase is explained by the exponential growth of the model size with respect to the number of actions, which is higher for the case with $2$ batteries.

\subsubsection{Failed runs.}
As visualized before in Fig.~\ref{fig:schematicActons}, a discrete action can lead to a successor state that violates one of the system constraints.
In total, $4.2$\% of the performed iterations for all cases combined resulted in a violation of either the frequency limits or the power line transmission capacity.
No significant differences in the percentage of failed runs is observed across the different cases.
Since these runs are incomplete and not representative for the simulated cases, they are not reported for further analysis.
Nevertheless, the number of failed runs provides an empirical indication of the adequacy of the scheduling limits for the reserve and flexibility deployment.
A high percentage of failed runs means that the ancillary service scheduling limits might be insufficient, and should be enlarged.

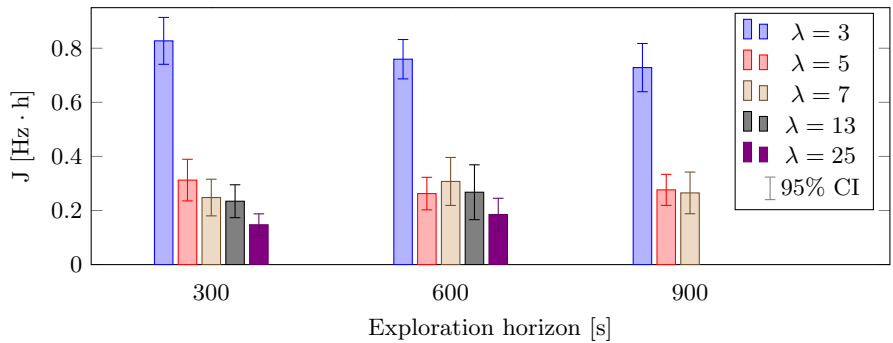
\begin{figure}[t!]
	\centering
	\scalebox{1}{\input{Figures/Results/Tikz_results_frequency}}

    \caption{Average total frequency deviations for the $3$-node case for different $\lambda$.}
	\label{fig:Results_cases_frequency}
\end{figure}

\subsubsection{Solution quality.}
The cost function of the MDP penalizes the sum of the absolute value of the frequency deviations in all grid nodes.
Therefore, the quality of the obtained solution can be evaluated by taking the integral of the total observed frequency deviation over the simulation horizon:
\begin{equation*}
    J = \int_{k_0}^{k_\text{end}} \big| \sum_{n=1}^{n_t} \omega_n(k) \big| dk,
\end{equation*}
where $k_0$ and $k_\text{end}$ cover the full 24-hour simulation time. The lower the value of $J$, the better the quality of the solution in terms of total frequency deviations.

A comparison of the value for $J$ between multiple cases on the $3$-node network from Fig.~\ref{fig:ModellingExampleGrid}, with different levels of $\lambda \in \{3, \ldots, 25\}$, is shown in Fig.~\ref{fig:Results_cases_frequency}.
Every bar shows the average results and the $95$\% confidence interval (CI) of the $100$ iterations performed for that case.
Due to infeasible run times, simulations for the cases with exploration horizon of $900$ seconds and $\lambda \geq 13$ were omitted.

We observe that (1) a \textit{longer exploration horizon does not improve the quality of the solution} in terms of frequency deviations significantly.
On the other hand, (2) \textit{increasing the number of discrete actions} in every dimension yields a significantly better solution quality.
This observation suggests that it is more beneficial to have a more fine-grained discretization of the continuous control space, than to invest in a longer optimization horizon.

\subsubsection{MDP model size.}
\begin{figure}[t!]
	\centering
	\hfill
	\scalebox{1}{\input{Figures/Results/Tikz_results_states}}
	\hfill
	\scalebox{1}{\input{Figures/Results/Tikz_results_actions}}
	\hfill${}$

	\caption{Average MDP states (left) and actions (right) for cases with $\lambda=5$, and varying numbers of nodes and batteries (plotted in log scale).}
	\label{fig:Results_cases_statesActions}
\end{figure}
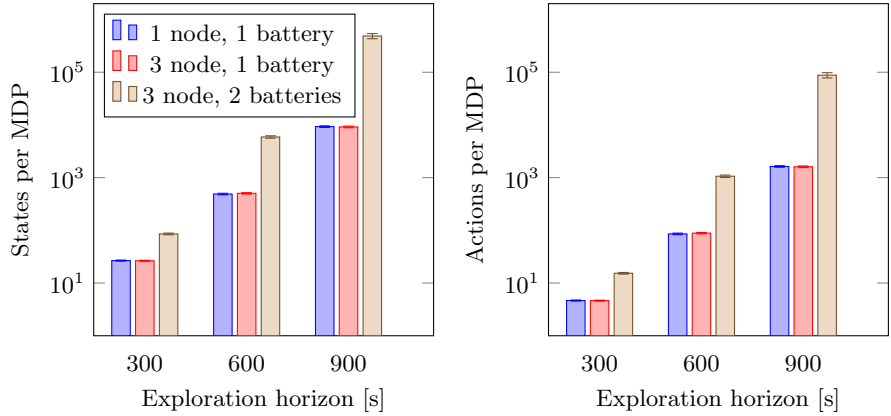

In Fig.~\ref{fig:Results_cases_statesActions}, the average number of states and actions per MDP are compared for different network configurations, all with $\lambda=5$.
All cases were simulated for $100$ runs, except for the case with $2$ batteries, which was only simulated for $1$ run, due to a too long run time.
We see that the number of states and actions is \textit{independent of the number of grid nodes}.
The intuition behind this is that an increased number of state features does not yield a larger MDP.
In fact, the branching factor of the applied exploration procedure only depends on the number of actions and successor states in the wind error DTMC, and is independent of the number of state features.
As expected, increasing the number of batteries to $2$ yields a significant increase in the number of MDP states and actions, especially for a longer exploration horizon, due to the additional dimension of the continuous control space.

\subsubsection{Frequency control and ancillary service deployment.}

\begin{figure*}[h!]
	\begin{subfigure}{0.516\textwidth}
		\scalebox{1}{\includegraphics[width=\linewidth]{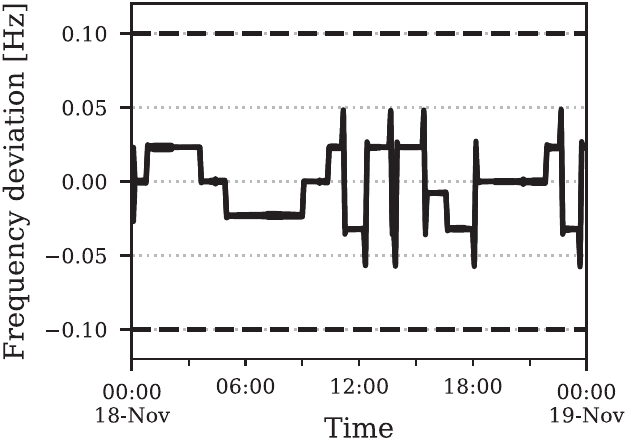}}
		\caption{Frequency deviations for $\lambda=3$.}
		\label{fig:Results_freqAnc_freq3}
	\end{subfigure}
	\hfill
	\begin{subfigure}{0.484\textwidth}
		\scalebox{1}{\includegraphics[width=\textwidth]{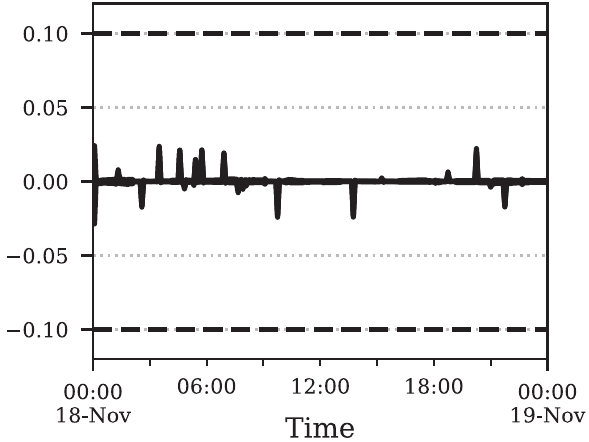}}
		\caption{Frequency deviations for $\lambda=25$.}
		\label{fig:Results_freqAnc_freq25}
	\end{subfigure}
	\hfill
	\hfill
	\begin{subfigure}{0.516\textwidth}
		\scalebox{1}{\includegraphics[width=\linewidth]{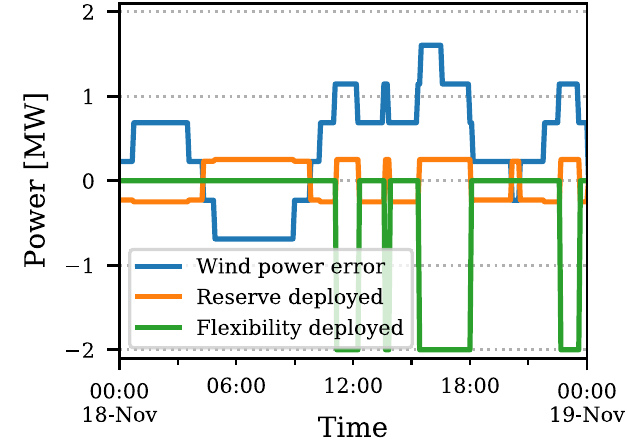}}
		\caption{Ancillary services for $\lambda=3$.}
		\label{fig:Results_freqAnc_ancillary3}
	\end{subfigure}
	\hfill
	\begin{subfigure}{0.484\textwidth}
		\scalebox{1}{\includegraphics[width=\textwidth]{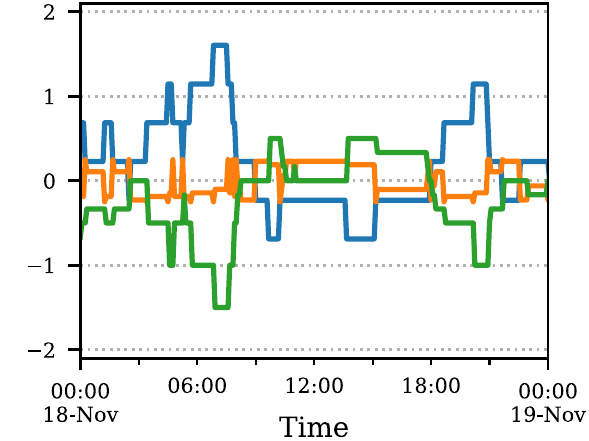}}
		\caption{Ancillary services for $\lambda=25$.}
		\label{fig:Results_freqAnc_ancillary25}
	\end{subfigure}
	\hfill
	\caption{Results for two runs for the 3-node case with 1 battery, for different $\lambda$.
	}
	\label{fig:Results_freqAnc}
\end{figure*}

Finally, in Fig.~\ref{fig:Results_freqAnc}, two example runs for the $3$-node network with a single battery and exploration horizon of $600$ seconds are shown.
Fig.~\ref{fig:Results_freqAnc_freq3} presents the frequency deviations for $\lambda=3$, while Fig.~\ref{fig:Results_freqAnc_freq25} shows the same for $\lambda=25$.
Similarly, Figs.~\ref{fig:Results_freqAnc_ancillary3} and \ref{fig:Results_freqAnc_ancillary25} show the deployment of reserves and flexibility for both cases.
Due to the uncertainty in the wind power forecast error, the results in Fig.~\ref{fig:Results_freqAnc} present only two possible trajectories, and repeating the experiment can lead to different results.

The observed frequency deviations are significantly lower for the case with more fine-grained discretization of the actions (\ie $\lambda = 25$), thus confirming the results also shown in Fig.~\ref{fig:Results_cases_frequency}.
This difference in the control precision is clearly depicted between Figs.~\ref{fig:Results_freqAnc_ancillary3} and \ref{fig:Results_freqAnc_ancillary25}.
The intuition behind this is that a fine-grained discretization allows for more precise control of the ancillary service deployment, which results in lower frequency deviations.

%% file: Figures/Results/Tikz_results_frequency.tex
\pgfplotsset{
  /pgfplots/error bar legend/.style={
    legend image code/.code={
        \draw[sharp plot,mark=-,mark repeat=2,mark phase=1,color=gray,##1]
        plot coordinates { (0.3cm, -0.15cm) (0.3cm,0cm) (0.3cm, 0.15cm) };%
        }}}

\begin{tikzpicture}
    \begin{axis}[
        width=\textwidth,
        height=5cm,
        ybar,
        bar width=7pt,
        xtick distance=1,
        xlabel=$\text{Exploration horizon [s]}$,
        ylabel=$\text{J [Hz}\cdot\text{h]}$,
        enlarge x limits={abs=0.5},
        xmin=150,
        xmax=1150,
        ymin=0,
        ymax=0.95,
        xtick={300,600,900},
        scaled ticks=false,
        xtick style={
            /pgfplots/major tick length=0pt,
        },
    ]

        \addplot+ [
            error bars/.cd,
                y dir=both,
                y explicit,
        ] coordinates {
            (300, 0.827246995) +- (300, 0.086667633)
            (600, 0.759484622) +- (600, 0.07278941)
            (900, 0.72843912) +- (900, 0.089132129)
        };
        
        \addplot+ [
            error bars/.cd,
                y dir=both,
                y explicit,
        ] coordinates {
            (300, 0.312477402) +- (300, 0.07695382)
            (600, 0.262571838) +- (600, 0.060136344)
            (900, 0.276344299) +- (900, 0.057405836)
        };
        
        \addplot+ [
            error bars/.cd,
                y dir=both,
                y explicit,
        ] coordinates {
            (300, 0.247979155) +- (300, 0.067773544)
            (600, 0.307539488) +- (600, 0.088748384)
            (900, 0.265107157) +- (900, 0.077078202)
        };
        
        \addplot+ [
            error bars/.cd,
                y dir=both,
                y explicit,
        ] coordinates {
            (300, 0.23438136) +- (300, 0.060815589)
            (600, 0.267655024) +- (600, 0.101366157)
        };
        
        \addplot+ [
            error bars/.cd,
                y dir=both,
                y explicit,
        ] coordinates {
            (300, 0.147340658) +- (300, 0.040362699)
            (600, 0.185207311) +- (600, 0.06003646)
        };
        
        \addplot [
        only marks,
        error bar legend,
        error bars/.cd,
        y dir=both, y explicit,
        error bar style={color=gray},
        ]
        coordinates {
          (-500, 0) +- (0, .5)
        };
        
        \legend{
            $\lambda = 3$,
            $\lambda = 5$,
            $\lambda = 7$,
            $\lambda = 13$,
            $\lambda = 25$,
            95\% CI
        }
    \end{axis}
\end{tikzpicture}

%% file: Figures/Results/Tikz_results_states.tex
\begin{tikzpicture}
    \begin{axis}[
        width=0.5\textwidth,
        height=6cm,
        ybar,
        ymode=log,
        bar width=7pt,
        xtick distance=1,
        xlabel=$\text{Exploration horizon [s]}$,
        ylabel=$\text{States per MDP}$,
        enlarge x limits={abs=0.5},
        xmin=150,
        xmax=1150,
        ymin=1,
        ymax=2000000,
        xtick={300,600,900},
        scaled ticks=false,
        xtick style={
            /pgfplots/major tick length=0pt,
        },
        legend pos  = north west,
    ]

        
        \addplot+ [
            error bars/.cd,
                y dir=both,
                y explicit,
        ] coordinates {
            (300, 26.6734) +- (300, 0.6085)
            (600, 488.6160) +- (600, 16.4074)
            (900, 9299.3044) +- (900, 307.3488)
        };
        
        \addplot+ [
            error bars/.cd,
                y dir=both,
                y explicit,
        ] coordinates {
            (300, 26.5085) +- (300, 0.5821)
            (600, 503.8912) +- (600, 15.1810)
            (900, 9175.1614) +- (900, 336.7776)
        };
        
        \addplot+ [
            error bars/.cd,
                y dir=both,
                y explicit,
        ] coordinates {
            (300, 85.5289) +- (300, 3.1801)
            (600, 5916.3453) +- (600, 310.0351)
            (900, 487137.4294) +- (900, 54377.2609)
        };
        
        
        \legend{
            {1 node, 1 battery},
            {3 node, 1 battery},
            {3 node, 2 batteries}
        }
    \end{axis}
\end{tikzpicture}

%% file: Figures/Results/Tikz_results_actions.tex
\begin{tikzpicture}
    \begin{axis}[
        width=0.5\textwidth,
        height=6cm,
        ybar,
        ymode=log,
        bar width=7pt,
        xtick distance=1,
        xlabel=$\text{Exploration horizon [s]}$,
        ylabel=$\text{Actions per MDP}$,
        enlarge x limits={abs=0.5},
        xmin=150,
        xmax=1150,
        ymin=1,
        ymax=2000000,
        xtick={300,600,900},
        scaled ticks=false,
        xtick style={
            /pgfplots/major tick length=0pt,
        },
        legend pos  = north west,
    ]
        
        \addplot+ [
            error bars/.cd,
                y dir=both,
                y explicit,
        ] coordinates {
            (300, 4.6621) +- (300, 0.1046)
            (600, 85.6775) +- (600, 2.9131)
            (900, 1630.7541) +- (900, 54.0094)
        };
        
        \addplot+ [
            error bars/.cd,
                y dir=both,
                y explicit,
        ] coordinates {
            (300, 4.6384) +- (300, 0.1019)
            (600, 88.5234) +- (600, 2.6824)
            (900, 1608.8734) +- (900, 59.2217)
        };
        
        \addplot+ [
            error bars/.cd,
                y dir=both,
                y explicit,
        ] coordinates {
            (300, 15.3341) +- (300, 0.5543)
            (600, 1064.6943) +- (600, 55.9397)
            (900, 87956.7647) +- (900, 10025.9583)
        };
        
    \end{axis}
\end{tikzpicture}

%% file: 5_conclusion_discussion.tex
\section{Concluding Remarks}

\label{sec:Conclusions}
We presented a novel method to solve the problem of short-term scheduling for flexibility-based ancillary services in power systems with uncertain wind power generation.
By modelling the problem as an MDP, we overcome the need for both sampling and linearization, as opposed to the continuous-state approaches used by most traditional power system analysis methods.
Our experiments show that our approach is feasible for power grids with different levels of complexity and under realistic operating conditions.
Furthermore, our results show it is more beneficial to have a more fine-grained discretization of the continuous control space, than to invest in a longer optimization horizon.
Since the size of the MDP grows exponentially with both the number of actions and the exploration horizon, making a trade-off between the two is necessary.

In the future, we will exploit the flexibility of our model to incorporate alternative grid configurations and  multiple sources of uncertainty, such as imperfect communication between assets in the grid, or demand uncertainty.
Moreover, instead of the batteries, other flexible assets can also be modeled, such as flexibility provided by the thermal inertia of large-scale buildings.

%% file: main.bbl
\begin{thebibliography}{10}
\providecommand{\url}[1]{\texttt{#1}}
\providecommand{\urlprefix}{URL }
\providecommand{\doi}[1]{https://doi.org/#1}

\bibitem{Aghaei2013}
Aghaei, J., Alizadeh, M.I.: {Demand response in smart electricity grids
  equipped with renewable energy sources: A review}. Renewable and Sustainable
  Energy Reviews  \textbf{18},  64--72 (2013)

\bibitem{DBLP:conf/ccece/Al-SaffarM19}
Al{-}Saffar, M., Mus{\'{\i}}lek, P.: Distributed optimal power flow for
  electric power systems with high penetration of distributed energy resources.
  In: {CCECE} 2019. pp.~1--5. {IEEE} (2019)

\bibitem{Badings2019buildings}
Badings, T.S.: MSc Thesis. Buildings-to-Grid Integration for Demand-Side
  Flexibility in Power Systems with Uncertain Generation. University of
  Groningen (2019)

\bibitem{Badings2019}
Badings, T.S., Rostampour, V., Scherpen, J.M.: {Distributed Building Energy
  Storage Units for Frequency Control Service in Power Systems}.
  IFAC-PapersOnLine  \textbf{52}(4),  228--233 (2019)

\bibitem{BK08}
Baier, C., Katoen, J.P.: Principles of Model Checking. MIT Press (2008)

\bibitem{Bertsch2016}
Bertsch, J., Hagspiel, S., Just, L.: {Congestion management in power systems:
  Long-term modeling framework and large-scale application}. Journal of
  Regulatory Economics  \textbf{50}(3),  290--327 (2016)

\bibitem{DBLP:books/cu/BV2014}
Boyd, S.P., Vandenberghe, L.: Convex Optimization. Cambridge University Press
  (2014)

\bibitem{DBLP:journals/tac/CalafioreC06}
Calafiore, G.C., Campi, M.C.: The scenario approach to robust control design.
  {IEEE} Trans. Autom. Control.  \textbf{51}(5),  742--753 (2006)

\bibitem{DBLP:journals/siamjo/CampiG08}
Campi, M.C., Garatti, S.: The exact feasibility of randomized solutions of
  uncertain convex programs. {SIAM} J. Optim.  \textbf{19}(3),  1211--1230
  (2008)

\bibitem{CA19}
Cauchi, N., Abate, A.: {\textbackslash}mathsf stochy : Automated verification
  and synthesis of stochastic processes. In: {TACAS} {(2)}. LNCS, vol. 11428,
  pp. 247--264. Springer (2019)

\bibitem{chertkov2017ensemble}
Chertkov, M., Chernyak, V.: Ensemble of thermostatically controlled loads:
  Statistical physics approach. Scientific reports  \textbf{7}(1), ~1--9 (2017)

\bibitem{Tao2020}
Ding, T., Zeng, Z., Bai, J., Qin, B., Yang, Y., Shahidehpour, M.: Optimal
  electric vehicle charging strategy with {M}arkov decision process and
  reinforcement learning technique. IEEE Transactions on Industry Applications
  \textbf{56}(5),  5811–5823 (2020)

\bibitem{ENTSOe}
{ENTSO-e}: {Transparency Platform - Generation Forecasts for Wind and Solar,
  Control area Germany} (2020)

\bibitem{Gerard2018}
Gerard, H., Rivero~Puente, E.I., Six, D.: {Coordination between transmission
  and distribution system operators in the electricity sector: A conceptual
  framework}. Utilities Policy  \textbf{50},  40--48 (2018)

\bibitem{grillo2016optimal}
{Grillo}, S., {Pievatolo}, A., {Tironi}, E.: Optimal storage scheduling using
  {M}arkov decision processes. IEEE Transactions on Sustainable Energy
  \textbf{7}(2),  755--764 (2016)

\bibitem{HHB12}
Hartmanns, A., Hermanns, H., Berrang, P.: A comparative analysis of
  decentralized power grid stabilization strategies. In: Winter Simulation
  Conference. pp. 158:1--158:13. {WSC} (2012)

\bibitem{Hemmati2017}
Hemmati, R., Saboori, H., Jirdehi, M.A.: {Stochastic planning and scheduling of
  energy storage systems for congestion management in electric power systems
  including renewable energy resources}. Energy  \textbf{133},  380--387 (2017)

\bibitem{Kempton2008}
Kempton, W., Udo, V., Huber, K., Komara, K., Letendre, S., Baker, S., Brunner,
  D., Pearre, N.: {A test of Vehicle-to-Grid (V2G) for Energy Storage and
  Frequency Regulation in the PJM System}. Results from an Industry-University
  Research Partnership  (2008)

\bibitem{Liu2018}
Liu, Y., Yu, N., Wang, W., Guan, X., Xu, Z., Dong, B., Liu, T.: {Coordinating
  the operations of smart buildings in smart grids}. Applied Energy
  \textbf{228}(July),  2510--2525 (2018)

\bibitem{Lymperopoulos2015}
Lymperopoulos, I., Qureshi, F.A., Nghiem, T., Khatir, A.A., Jones, C.N.:
  {Providing ancillary service with commercial buildings: The Swiss
  perspective}. IFAC-PapersOnLine  \textbf{28}(8),  6--13 (2015)

\bibitem{MacDougall2013}
MacDougall, P., Roossien, B., Warmer, C., Kok, K.: {Quantifying flexibility for
  smart grid services}. In: 2013 IEEE Power Energy Society General Meeting.
  pp.~1--5. IEEE (2013)

\bibitem{Machowski2006PowerControl}
Machowski, J., Dong, Z.Y., Member, S., Zhang, P.: {Power System Dynamics:
  Stability and Control}. Wiley (2006)

\bibitem{DBLP:journals/tac/MargellosGL14}
Margellos, K., Goulart, P., Lygeros, J.: On the road between robust
  optimization and the scenario approach for chance constrained optimization
  problems. {IEEE} Trans. Autom. Control.  \textbf{59}(8),  2258--2263 (2014)

\bibitem{Margellos2012}
Margellos, K., Haring, T., Hokayem, P., Schubiger, M., Lygeros, J., Andersson,
  G.: {A Robust Reserve Scheduling Technique for Power Systems with High Wind
  Penetration}. Proceedings of PMAPS pp. 870--875 (2012)

\bibitem{DBLP:journals/mp/MurtyK87}
Murty, K.G., Kabadi, S.N.: Some np-complete problems in quadratic and nonlinear
  programming. Math. Program.  \textbf{39}(2),  117--129 (1987)

\bibitem{NGESO2020}
{NG ESO}: Optional downward flexibility management (odfm) service documents.
  National Grid ESO  (2020)

\bibitem{DBLP:journals/tase/NguyenSB17}
Nguyen, D.B., Scherpen, J.M.A., Bliek, F.: Distributed optimal control of smart
  electricity grids with congestion management. {IEEE} Trans Autom. Sci. Eng.
  \textbf{14}(2),  494--504 (2017)

\bibitem{Papaefthymiou2008}
Papaefthymiou, G., Kl{\"{o}}ckl, B.: {MCMC for wind power simulation}. IEEE
  Transactions on Energy Conversion  \textbf{23}(1),  234--240 (2008)

\bibitem{DBLP:conf/qest/PeruffoGPA19}
Peruffo, A., Guiu, E., Panciatici, P., Abate, A.: Safety guarantees for the
  electricity grid with significant renewables generation. In: {QEST}. LNCS,
  vol. 11785, pp. 332--349. Springer (2019)

\bibitem{Pillay2015}
Pillay, A., Prabhakar~Karthikeyan, S., Kothari, D.P.: {Congestion management in
  power systems - A review}. International Journal of Electrical Power and
  Energy Systems  \textbf{70},  83--90 (2015)

\bibitem{DBLP:books/wi/Puterman94}
Puterman, M.L.: {M}arkov Decision Processes: Discrete Stochastic Dynamic
  Programming. Wiley Series in Probability and Statistics, Wiley (1994)

\bibitem{DBLP:journals/tsg/RazmaraBSPR18}
Razmara, M., Bharati, G.R., Shahbakhti, M., Paudyal, S., Robinett, R.D.:
  Bilevel optimization framework for smart building-to-grid systems. {IEEE}
  Trans. Smart Grid  \textbf{9}(2),  582--593 (2018)

\bibitem{DBLP:conf/cdc/RostampourBS19}
Rostampour, V., Badings, T.S., Scherpen, J.M.A.: Buildings-to-grid integration
  with high wind power penetration. In: {CDC}. pp. 2976--2981. {IEEE} (2019)

\bibitem{RostampourBadings2020}
Rostampour, V., Badings, T.S., Scherpen, J.M.A.: Demand flexibility management
  for buildings-to-grid integration with uncertain generation. Energies
  \textbf{13}(24) (2020)

\bibitem{rostampour2017distributedTPS}
Rostampour, V., Ter~Haar, O., Keviczky, T.: Distributed stochastic reserve
  scheduling in ac power systems with uncertain generation. IEEE Transactions
  on Power Systems  \textbf{34}(2),  1005--1020 (2018)

\bibitem{Sincovec1981}
Sincovec, R.F., Erisman, A.M., Yip, E.L., Epton, M.A.: {Analysis of Descriptor
  Systems Using Numerical Algorithms}. {IEEE} Trans. Autom. Control.
  \textbf{26}(1),  139--147 (1981)

\bibitem{DBLP:journals/tcst/SoudjaniA15}
Soudjani, S.E.Z., Abate, A.: Aggregation and control of populations of
  thermostatically controlled loads by formal abstractions. {IEEE} Trans.
  Control. Syst. Technol.  \textbf{23}(3),  975--990 (2015)

\bibitem{SGA15}
Soudjani, S.E.Z., Gevaerts, C., Abate, A.: {FAUST}\({}^{\mbox{ 2}}\) : Formal
  abstractions of uncountable-state stochastic processes. In: {TACAS}. LNCS,
  vol.~9035, pp. 272--286. Springer (2015)

\bibitem{DBLP:journals/tsg/TahaGDPL19}
Taha, A.F., Gatsis, N., Dong, B., Pipri, A., Li, Z.: Buildings-to-grid
  integration framework. {IEEE} Trans. Smart Grid  \textbf{10}(2),  1237--1249
  (2019)

\bibitem{DBLP:conf/cdc/TripBP14}
Trip, S., B{\"{u}}rger, M., Persis, C.D.: An internal model approach to
  frequency regulation in inverter-based microgrids with time-varying voltages.
  In: {CDC}. pp. 223--228. {IEEE} (2014)

\bibitem{Wang2011}
Wang, J., Liu, C., Ton, D., Zhou, Y., Kim, J., Vyas, A.: {Impact of plug-in
  hybrid electric vehicles on power systems with demand response and wind
  power}. Energy Policy  \textbf{39}(7),  4016--4021 (2011)

\end{thebibliography}
